\documentstyle[11pt]{article}
\begin{document}
 
\pagestyle{empty}
\baselineskip=20.0pt

\centerline{Properties of Nuclear and Neutron Matter and Thermodynamic Consistency}
\centerline{in a Nonlinear Mean-field Approximation}

\vspace{1.5cm}

\centerline{Hiroshi Uechi\footnote{E-mail: uechi@utc.osaka-gu.ac.jp}}

\vspace{1.0cm}

\centerline{Osaka Gakuin Junior College}
\centerline{2-37-1  Kishibe-minami,  Suita,  Osaka  564-8511 \quad JAPAN}

\vspace{1.5cm}

\centerline{\large{Abstract}}

Properties of nuclear and neutron matter are discussed in a nonlinear $\sigma$-$\omega$-$\rho$ mean-field approximation with self-interactions
and mixing-interactions of mesons and baryons.  The nonlinear interactions are renormalized by employing the theory of conserving
approximations, which results in a thermodynamically consistent approximation that maintains Hugenholtz-Van Hove theorem and Landau's requirement
of quasiparticles.  The approximation is equivalent to the Hartree approximation with {\it effective masses} and {\it effective coupling constants}
of baryons and mesons.  The effective masses and coupling constants are naturally required by self-consistency of the theory of conserving
approximations.   The approximation is applied to
nuclear and neutron matter, which suggests that the lower bound of nuclear compressibility $K \sim 180$ MeV (with the symmetry energy
$a_4 = 35.0$ MeV) be required to be consistent with properties of nuclear matter and the maximum masses of observed hadronic neutron stars
($M_{max} \ge 2.00$ $M_{\odot}$).  The values of the compressibility, symmetry energy together with effective masses and coupling constants of
baryons and mesons will be important constraints to examine models of nuclear and neutron matter.  The accumulating data and accurate
measurements of observables in high density and energy region will supply significant information in order to testify theoretical consistency of
nuclear models.

\newpage

\makeatletter
\renewcommand{\theequation}{%
\thesection.\arabic{equation}}
\@addtoreset{equation}{section}
\makeatother

\baselineskip=20.0pt
\setcounter{page}{1}
\newcommand{\bkappa}{\mbox{\boldmath $\kappa$}}
\newcommand{\btau}{\mbox{\boldmath $\tau$}}
\newcommand\kfermi{k_{\scriptscriptstyle F}}
\newcommand\Mstar{M^{\ast}}
\newcommand\kstar{k^{\ast}}
\newcommand\rhoB{\rho_{\scriptscriptstyle B}}
\newcommand\szero{\scriptscriptstyle (0)}
\newcommand{\bftau}{\mbox{\boldmath $\tau$}}

\pagestyle{plain}

\section{Introduction}

The linear $\sigma$-$\omega$ mean-field approximation has been extensively applied to finite nuclei, nucleon-nucleon
interactions as well as to equations of state of nuclear matter and properties of neutron stars$^{\cite{WAL,SER}}$.  The
structure of the linear mean-field approximation is extended to nonlinear mean-field models which have also been extensively
applied to properties of nuclear and neutron matter$^{\cite{MUL,GLE}}$.  Neutron stars have attracted much interest
in nuclear and high energy physics with a large body of accumulating experimental evidence.  The observed
masses of hadronic neutron stars are above 1.3 $M_{\odot}$ (the solar mass: $M_{\odot} \sim 1.989 \times 10^{30}\
{\rm kg}$), and the maximum masses of neutron stars are expected to be below $2.5\ M_{\odot}$$^{\cite{AKM}}$,
or $2.2 \pm 0.2 M_{\odot}$$^{\cite{LAT}}$.   Since a neutron star is mostly composed of hadrons, the properties of a neutron star may be
calculated with the self-consistent equation of state for the many-body system of hadrons.  We will employ a nonlinear mean-field model to
investigate relations between properties of nuclear matter and neutron stars constrained by the maximum masses of neutron stars:
$M_{max} = 2.50\ M_{\odot}$, and $2.00\ M_{\odot}$.

Approximations for infinite nuclear and neutron matter are required to maintain self-consistent conditions: the Hugenholtz-Van Hove
theorem$^{\cite{HUG}}$ (HV theorem), the virial theorem$^{\cite{WEE}}$ and the theory of conserving approximations$^{\cite{BAY}}$.  The
self-consistent conditions are essential to calculate explicitly microscopic and macroscopic physical quantities and denoted
as {\it thermodynamic consistency}$^{\cite{BAY,UEC}}$ to an approximation.  Serot and Walecka's approach is extended by maintaining
thermodynamic consistency in the current nonlinear model, and the extension leads to the concept of {\it effective masses of mesons}
and {\it effective coupling constants}.  The mean-field approximation with nonlinear interactions of baryons and mesons will become
a conserving approximation when effective masses of mesons and effective vertex coupling constants are self-consistently defined.  This results
in an important conclusion that the nonlinear interactions of mesons should be examined as density-dependent, or energy-dependent effective
masses and coupling constants.

The nonlinear mean-field model contains several adjustable parameters coming from self-interactions and mixing-interactions of mesons and
baryons.  Typically, the treatment of quantum vacuum is neglected entirely that it may be difficult to study nuclear systems at short
distances, or high-energy regions$^{\cite{WAL,SER}}$.  However, the current nonlinear mean-field approximation constructed with thermodynamic
consistency is a self-consistent approximation which constrains nonlinear interactions and adjustable parameters in terms of physical conservation
laws.  In this sense, the general features obtained in the calculation may compensate for the deficiency and remain valid as model-independent results.

The nonlinear mean-field approximation will be proved equivalent to Hartree approximation with effective masses and effective vertex coupling
constants.  The results indicate that nonlinear interactions have a significant effect on properties of nuclear and neutron matter; however, since
nonlinear interactions are strongly confined by self-consistency, their contributions to physical quantities are obtained in a balanced and restricted
way.  Hence, the existence and strength of nonlinear interactions of hadrons could be examined if effective masses of mesons are measured
from the nuclear experimental data.  The renormalized effective masses and vertex coupling constants are derived explicitly, and physical
contributions of nonlinear interactions to properties of nuclear and neutron matter are checked quantitatively.  The analysis
of nonlinear interactions by way of effective masses and coupling constants is one of the purposes of the current paper.  The self-consistency
of meson-propagators with effective masses of mesons and effective coupling constants will be explained in detail in the sec.~3 and sec.~4.

In order to simulate complicated interactions and retardation effects, the nonlinear $\sigma$ mean-field$^{\cite{BOG}}$ and the
nonlinear $\sigma$-$\omega$-$\rho$ mean-field models$^{\cite{MUL}}$ have been discussed.  By adjusting coupling constants of
nonlinear self-interactions of mesons, it seems that one could obtain softer equations of state as one wishes$^{\cite{MUL}}$
in order to simulate nuclear properties.   However, since the nonlinear interactions can be renormalized as
{\it effective masses and coupling constants}, contributions to physical quantities will be restricted by self-consistency; hence, there appear
a possibility to check the validity of nonlinear interactions by experimental data of nuclear matter and neurton stars.  Furthermore, if
nonlinear mixing interactions are introduced, coupled equations of motion for mesons and baryons will confine the values of coupling constants
more strictly.  Therefore, coupling constants of nonlinear interactions are not free parameters at all.   Coupled equations of motion and
self-consistency as a conserving approximation will restrict the strength of nonlinear interactions.  This is an important consequence derived from
the current conserving mean-field approximation.

The properties of neutron stars, such as the mass, moment of inertia and radius of neutron stars, are calculated by
employing the equation of state in the nonlinear mean-field approximation.  The coupling constants are confined within certain
values in order to produce saturation properties and the symmetry energy, $a_4 = 35.0$ MeV$^{\cite{WAL}}$, of nuclear matter
and masses of neutron stars.  We have calculated properties of nuclear matter and neutron stars (nuclear compressibility,
symmetry energy, effective masses of nucleons and mesons, the maximum mass of neutron stars) and investigated the minimum
value of nuclear compressibility, $K$, to the value of the supposed maximum mass of neutron stars, $M_{max} = 2.50\ M_{\odot}$ and
2.00 $M_{\odot}$ .   The nonlinear mean-field approximation suggests that a lower bound of compressibility
$K > 340$ MeV should be required to be compatible with the maximum value, $M_{star} > 2.50 \ M_{\odot}$; and
$K > 180$ MeV compatible with $M_{star} > 2.00 \ M_{\odot}$.  

It is difficult to reproduce $K$ and $a_4$, $M_{max}$, simultaneously by adjusting the values of coupling constants.  The conditions of
thermodynamic consistency and coupled equations of motion for mesons and baryons strictly restrict contributions of nonlinear
interactions.  Consequently, the values of nonlinear coupling constants are confined in a narrow region.  Therefore,
nonlinear interactions will generate certain fixed experimental values, and the validity of nonlinear interactions and relations
between properties of nuclear matter and neutron stars may be examined by comparing with experimental data of nuclear matter and neutron
stars.   These results are explained in the sec.~5.  Conclusions and remarks on the self-consistent nonlinear mean-field approximation are in the sec.~6.

\section{The nonlinear $\sigma$-$\omega$-$\rho$ mean-field lagrangian with mixing and self-interactions of mesons
and baryons}

The mean-field lagrangian of the nonlinear $\sigma$-$\omega$-$\rho$ self-interactions, mixing-interactions,
$\sigma \omega N$ and $\sigma \rho N$ couplings of mesons and nucleons is defined by,
\begin{eqnarray}
{\cal L}_{\scriptscriptstyle MFT} = &&\hspace{-0.6cm} {\bar \psi} \Bigl[ i \gamma_{\mu} \partial^{\mu} -
g_{\omega} \gamma_0 V_0 - g_{\sigma \omega N} \phi_0 \gamma_0 V_0 - \frac{g_{\rho}}{2} \gamma_0 \tau_3 R_0 -
g_{\sigma \rho N}\phi_0 \gamma_0 \tau_3 R_0 - (M - g_{\sigma} \phi_0 ) \Bigr] \psi \nonumber \\
&&- \frac{1}{2} m_{\sigma}^2 \phi_0^2 - \frac{g_{\sigma 3}}{3!} \phi_0^3 - \frac{g_{\sigma 4}}{4!} \phi_0^4
+ \frac{1}{2} m_{\omega}^2 V_0^2 + \frac{g_{\omega 4}}{4!} V_0^4 + \frac{g_{\sigma \omega}}{4} \phi_0^2 V_0^2 \nonumber \\
&&+ \frac{1}{2} m_{\rho}^2 R_0^2 + \frac{g_{\rho 4}}{4!} R_0^4 
+ \frac{g_{\sigma \rho}}{4} \phi_0^2 R_0^2 + \frac{g_{\omega \rho}}{4} V_0^2 R_0^2 \label{eqn:MFL} \ ,
\end{eqnarray}
where the meson-field operators are replaced by expectation values in the ground state: $\phi_0$ for the $\sigma$-field, $V_0$ for the
vector-isoscalar $\omega$-meson, $V_{\mu} V^{\mu} = V_0^2 - {\bf V}^2$, $(\mu = 0, 1, 2, 3)$; the neutral $\rho$-meson mean-field, $R_0$,
is chosen for $\tau_3$-direction in isospin space.   The density-dependent effective vertex interactions of $g_{\sigma \omega N}$ and
$g_{\sigma \rho N}$ terms are defined by way of the scalar mean-field, $\phi_0$.  Let us notice that the replacement of the meson quantum fields
with classical fields means that the nonlinear mean-field lagrangian should be understood as an effective model to simulate complex many-body
interactions as simple as possible in the mean-field approximation.

The nonlinear model is motivated by preserving the structure of Serot and Walecka's linear $\sigma$-$\omega$
mean-field approximation$^{\cite{WAL}}$ as explicit as possible,
Lorentz-invariance and renormalizability, thermodynamic consistency: the Hugenholtz-Van Hove theorem$^{\cite{HUG}}$, the virial theorem$^{\cite{WEE}}$,
and conditions of conserving approximations$^{\cite{BAY,UEC}}$.  The concept of {\it effective masses} and {\it effective coupling constants} are naturally generated by self-consistent conditions and will be proved by the requirements of
theorems and conditions of conserving approximations.

The masses in~($\ref{eqn:MFL}$) are: $M = 939$ MeV, $m_{\sigma} = 550$ MeV, $m_{\omega} = 783$ MeV and $m_{\rho} = 770$ MeV, in order to
compare the effects of nonlinear interactions with those of the linear $\sigma$-$\omega$ approximation discussed by Serot and
Walecka$^{\cite{WAL}}$.  The coupling constants may be supposed to be free parameters at the outset, however, they are constrained by saturation
properties: the binding energy $-15.75$ MeV at $\kfermi = 1.30$ fm$^{-1}$ with the symmetry energy, $a_4 = 35.0$ MeV, and in addition, by
investigating the lower bound of nuclear compressibility, $K$, which corresponds to maximum masses of neutron stars.  The maximum masses of
observed neutron stars are chosen in the current analysis as $M_{max} = 2.50$ M$_{\odot}$ and 2.00 M$_{\odot}$$^{\cite{AKM,LAT}}$, in order to
examine contributions of self-consistently constrained nonlinear interactions in the level of Hartree approximation.  

The equation of motion for baryons is given by,
\begin{equation}
\Bigl[ (i\gamma_{\mu} \partial^{\mu} - g_{\omega}^{\ast} \gamma_0 V_0 - \frac{g_{\rho}^{\ast}}{2} \gamma_0 \tau_3 R_0)
- (M - g_{\sigma} \phi_0) \Bigr] \psi = 0 \label{eqn:EQB} \ ,
\end{equation}
where $g_{\omega}^{\ast} = g_{\omega} + g_{\sigma \omega N} \phi_0$ and $g_{\rho}^{\ast} = g_{\rho} + 2 g_{\sigma \rho N} \phi_0$
are {\it effective coupling constants} for $\omega$ and $\rho$ mesons, respectively.   The equations of motion for mesons in the
mean-field approach are given by,
\begin{eqnarray}
m_{\sigma}^2 \phi_0 + \frac{g_{\sigma 3}}{2!} \phi_0^2 + \frac{g_{\sigma 4}}{3!} \phi_0^3 - \frac{g_{\sigma \omega}}{2} V_0^2 \phi_0
- \frac{g_{\sigma \rho}}{2} R_0^2 \phi_0
=&&\hspace{-0.6cm} g_{\sigma} \langle {\bar \psi} \psi \rangle - g_{\sigma \omega N} V_0 \langle \psi^{\dagger}\psi \rangle
- g_{\sigma \rho N} R_0 \langle \psi^{\dagger}\tau_3 \psi \rangle \ , \nonumber \\
m_{\omega}^2 V_0 + \frac{g_{\omega 4}}{3!} V_0^3 + \frac{g_{\sigma \omega}}{2} \phi_0^2 V_0 + \frac{g_{\omega \rho}}{2} R_0^2 V_0
=&&\hspace{-0.6cm} g_{\omega}^{\ast} \langle \psi^{\dagger}\psi \rangle = g_{\omega}^{\ast} \rhoB  \ , \nonumber \\
m_{\rho}^2 R_0 + \frac{g_{\rho 4}}{3!} R_0^3 + \frac{g_{\sigma \rho}}{2} \phi_0^2 R_0 + \frac{g_{\omega \rho}}{2} V_0^2 R_0
=&&\hspace{-0.6cm} \frac{g_{\rho}^{\ast}}{2} \langle \psi^{\dagger}\tau_3 \psi \rangle = \frac{g_{\rho}^{\ast}}{2} \rho_3 \label{eqn:EMM} \ .
\end{eqnarray}
The scalar density, $\rho_s$, is defined by
\begin{equation}
\rho_s = \langle {\bar \psi} \psi \rangle = \frac{\zeta}{(2\pi)^3} \int^{\kfermi}_0 \! d^3k \frac{\Mstar}{E^{\ast} (k)} \ ,
\end{equation}
where $\Mstar = M - g_{\sigma} \phi_0$.  The baryon density will be denoted as $\displaystyle \rhoB = \langle \psi^{\dagger}\psi \rangle =
\frac{\zeta \kfermi^3}{6\pi^2}$ ($\zeta = 4$ for nuclear matter; $\zeta = 2$ for neutron matter), and
$\rho_3 = \langle \psi^{\dagger}\tau_3 \psi \rangle = (k^3_{F_p} - k^3_{F_n})/3\pi^2$; the Fermi-momentum $k_{F_p}$
is for proton and $k_{F_n}$ for neutron.  In the linear $\sigma$-$\omega$ mean-field approximation, the scalar source is only generated
by the scalar density, $\rho_s$.  However, one should note that the nonlinear mixing interactions, such as ${\bar \psi} \gamma_0 V_0 \psi$ and
${\bar \psi} \gamma_0 \tau_3 R_0 \psi$, generate additional source terms to the equation of motion for the scalar field, $\phi_0$.  The equations
of motion for $\omega$ and $\rho$ mesons are formally obtained by replacing $g_{\omega}$ and $g_{\rho}$ in the linear mean-field approximation
to $g_{\omega}^{\ast}$ and $g_{\rho}^{\ast}$.

{\it The theory of conserving approximations} requires that the equations of motion be derived as the functional
derivative of energy density with respect to self-energies$^{\cite{UEC}}$ (or equivalently, with respect to mean-fields: $\phi_0$, $V_0$, $R_0$),
and moreover, it requires that the equations of motion expressed in terms of self-energies be derived by Feynman diagrams by applying propagators of baryons and
mesons.  It is essential for the theory of infinite matter to prove the equivalence of both calculations.  The propagators of baryons
and mesons are defined in the sec.~3; the functional derivative of energy density with respect to mean-fields of mesons and the equivalence of
both approach will be proved in the sec.~4.

\section{The nonlinear $\sigma$-$\omega$-$\rho$ Hartree approximation}

The approximation-scheme of Serot and Walecka's linear $\sigma$-$\omega$ approximation is maintained by employing {\it effective coupling constants}
and {\it effective masses}, and it is readily proved that the current approximation is a self-consistent, relativistic Hartree approximation as
discussed in the linear $\sigma$-$\omega$ approximation.

The Green's function for baryons satisfies Dyson-Schwinger equation, and the formal solution to the Green's function is:
\begin{eqnarray}
[ G(k) ]^{-1} =\hspace{-0.6cm}&& \gamma_{\mu} (k^{\mu} + \Sigma^{\mu}(k)) - (M + \Sigma^s(k)) \nonumber \ ,\\
=\hspace{-0.6cm}&& \gamma_0 (k^0 + \Sigma^0(k)) - {\bf \gamma}\cdot {\bf k}(1 + \Sigma^v (k)) - (M + \Sigma^s(k)) \label{eqn:GK} \ ,
\end{eqnarray}
where the spatial part of the self-energy, $\Sigma^i (k)$, is expressed as $\Sigma^i (k) = k^i \Sigma^v (k)$
$(i = 1, 2, 3)$; however, in the level of the mean-field approximation, the self-energy of the spatial part will vanish
because of the spatial homogeneity, resulting in $\Sigma^v (k) = 0$.  As explained in the sec.~1, the quantum vacuum is entirely neglected, and
the Green's function for the Fermi-sea particles in the rest frame of nuclear matter is used:
\begin{equation}
G_{\scriptscriptstyle D}(k) = (\gamma^{\mu} k_{\mu}^{\ast} + \Mstar (k)) {i\pi \over E^{\ast}(k)} \delta (k^0 - E(k))
\theta (\kfermi - |{\bf k}|) \ , \label{eqn:gdd}
\end{equation}
where $k^{\ast 0} = k^0 + \Sigma^0$.  The effective nucleon mass is $\Mstar (\kfermi) = M - g_{\sigma} \phi_0 = M + \Sigma^s (\kfermi)$, and
the baryon single particle energy is $E(k) = E^{\ast}(k) - \Sigma^0 (k)$, with $E^{\ast}(k) = ({\bf k}^2 + M^{\ast 2})^{1/2}$.

The mean-fields of mesons and self-energies of baryons are specifically associated by way of the formal solution to Green's
function~($\ref{eqn:GK}$) and the equation of motion for baryons~($\ref{eqn:EQB}$); the $\omega$-meson contribution to the baryon
self-energy is denoted by $\Sigma_{\omega}^0 = -g_{\omega}^{\ast} V_0$, and $\rho$-meson contributions to the self-energies for proton
and neutron are denoted as $\Sigma_{\rho p}^0 = -(g_{\rho}^{\ast}/2) R_0$ and $\Sigma_{\rho n}^0 = (g_{\rho}^{\ast}/2) R_0$, respectively.

According to the linear Hartree approximation$^{\cite{WAL}}$, we will define the free, $\sigma$, $\omega$ and $\rho$, meson-propagators,
$D_{\sigma}^{\szero} (k)$, $D_{\omega}^{{\szero} \mu \nu} (k)$, and $D^{{\szero} \mu \nu}_{\rho \ a b} (k)$, for the nonlinear mean-field approximation as:
\begin{equation}
D_{\sigma}^{\szero} (k) = \frac{1}{k_{\lambda}^2 - m_{\sigma}^{\ast 2} + i\varepsilon} \ , \qquad
D_{\omega}^{{\szero} \mu \nu} (k) = \frac{-g^{\mu \nu}}{k_{\lambda}^2 - m_{\omega}^{\ast 2} + i\varepsilon} \ , \label{eqn:SVP}
\end{equation}
and
\begin{equation}
D^{{\szero} \mu \nu}_{\rho \ a b} (k) = \frac{-g^{\mu \nu}\delta_{a b}}{k_{\lambda}^2 - m_{\rho}^{\ast 2} + i\varepsilon} \ ,
\end{equation}
where $m_{\sigma}^{\ast}$, $m_{\omega}^{\ast}$ and $m_{\rho}^{\ast}$ are effective masses of $\sigma$, $\omega$, $\rho$ mesons,
respectively.  The effective masses are density-dependent constants, and therefore, they are determined when energy density,
pressure and self-energies are self-consistently determined.

The self-energies are calculated by applying $G_{\scriptscriptstyle D} (k)$, $D_{\sigma}^{\szero} (k)$, $D_{\omega}^{{\szero} \mu \nu} (k)$
and $D^{{\szero} \mu \nu}_{\rho \ a b} (k)$ upon (tadpole) Feynman diagrams as shown in Fig.~1a and Fig.~1b.  The interaction-lines in Fig.~1a and
Fig.~1b are given by effective masses of mesons which should be self-consistently defined.  The vertex factor of baryon and $\rho$-meson coupling
is $-i (g_{\rho}^{\ast}/2) \tau_3 \gamma^{\mu}$.  Let us notice that the scalar self-energy, $\Sigma^s$, is generated by three density sources
in the nonlinear model.  The scalar self-energy given by the Feynman diagram in Fig.~1a is
\begin{equation}
\Sigma^s = i \frac{g_{\sigma}}{m_{\sigma}^{\ast 2}} \int\! \frac{d^4 q}{(2\pi)^4} {\rm Tr} \Bigl\{ (g_{\sigma}-g_{\sigma\omega N}
V_0\gamma^0 -g_{\sigma\rho N}R_0\gamma^0\tau_3) G_{\scriptscriptstyle D} (q) \Bigr\} = -\frac{g_{\sigma}^2}{m_{\sigma}^{\ast 2}}
\rho_s^{\ast} \ , \label{eqn:SS}
\end{equation}
where $\rho_s^{\ast}$ is the modified scalar density: $\displaystyle \rho_s^{\ast} = (\rho_s - \frac{g_{\sigma\omega N}}{g_{\sigma}}
V_0 \rhoB - \frac{g_{\sigma\rho N}}{g_{\sigma}} R_0 \rho_3)$.  The $\omega$-meson and $\rho$-meson contributions to the self-energy are given by
\begin{equation}
\displaystyle \Sigma_{\omega}^{\mu} = i \frac{g_{\omega}^{\ast 2}}{m_{\omega}^{\ast 2}} \int\! \frac{d^4 q}{(2\pi)^4} {\rm Tr}
\Bigl( \gamma^{\mu} G_{\scriptscriptstyle D} (q) \Bigr) = - \frac{g_{\omega}^{\ast 2}}{m_{\omega}^{\ast 2}} \rhoB \delta_{\mu, 0} \ , \label{eqn:OS}
\end{equation}
and
\begin{equation}
\Sigma^{\mu}_{\rho(^p_n)} = \pm i \frac{g_{\rho}^{\ast 2}}{4 m_{\rho}^{\ast 2}} \int\! \frac{d^4 q}{(2\pi)^4} {\rm Tr}
\Bigl( \tau^3 \gamma^{\mu} G_{\scriptscriptstyle D} (q) \Bigr) = \mp \frac{g_{\rho}^{\ast 2}}{4 m^{\ast 2}_{\rho}} \rho_3 \delta_{\mu, 0} \ , \label{eqn:0R}
\end{equation}
where $\tau_3$ is the isospin matrix.  Self-energies, $(\ref{eqn:SS}) \sim (\ref{eqn:0R})$, are formally identical with the equations in
the linear mean-field approximation$^{\cite{WAL}}$.  As the effective mass of baryons is defined so that Dirac equation of motion is preserved, the
effective masses of mesons are defined so that they are formally identical with the equations of motion of mesons in the linear mean-field approximation.

The self-energies of mesons, $(\ref{eqn:SS}) \sim (\ref{eqn:0R}$), derived from Feynman diagrams are consistent with equations of motion of
mesons $(\ref{eqn:EMM})$.  Since the self-energies and mean-fields are related with $\Sigma^s (\kfermi) = -g_{\sigma} \phi_0$, $\Sigma^0_{\omega} (\kfermi)
= - g_{\omega}^{\ast} V_0$, and $\Sigma^0_{\rho(^p_n)} (\kfermi) = \mp (g_{\rho}^{\ast}/2) R_0$ ($(-)$ for proton and $(+)$ for neutron),
the self-energies from~($\ref{eqn:SS}$) to~($\ref{eqn:0R}$) must be the same as the meson equations of motion, ($\ref{eqn:EMM}$),
respectively.  Therefore, by comparing the self-energies from~($\ref{eqn:SS}$) to~($\ref{eqn:0R}$) and equations of motion for mesons
($\ref{eqn:EMM}$), one can examine that the current nonlinear mean-field approximation will be self-consistent, only if effective masses,
$m_{\sigma}^{\ast}$, $m_{\rho}^{\ast}$ and $m_{\omega}^{\ast}$, are given by,
\begin{eqnarray}
m_{\sigma}^{\ast 2} =&&\hspace{-0.6cm} m_{\sigma}^2 \Bigl( 1 + \frac{g_{\sigma 3}}{2 m_{\sigma}^2} \phi_0
+ \frac{g_{\sigma 4}}{3! m_{\sigma}^2} \phi_0^2 - \frac{g_{\sigma \omega}}{2 m_{\sigma}^2} V_0^2
+ \frac{\zeta - 4}{\zeta} \frac{g_{\sigma \rho}}{2 m_{\sigma}^2} R_0^2 \Bigr) \ , \nonumber \\
m_{\omega}^{\ast 2} =&&\hspace{-0.6cm} m_{\omega}^2 \Bigl( 1 + \frac{g_{\omega 4}}{3! m_{\omega}^2} V_0^2
+ \frac{g_{\sigma \omega}}{2 m_{\omega}^2} \phi_0^2 - \frac{\zeta - 4}{\zeta} \frac{g_{\omega \rho}}{2 m_{\omega}^2} R_0^2 \Bigr) \ ,
\nonumber \\
m_{\rho}^{\ast 2} =&&\hspace{-0.6cm} m_{\rho}^2 \Bigl( 1 - \frac{\zeta - 4}{\zeta} (\frac{g_{\rho 4}}{3! m_{\rho}^2} R_0^2
+ \frac{g_{\sigma \rho}}{2 m_{\rho}^2} \phi_0^2 + \frac{g_{\omega \rho}}{2 m_{\rho}^2} V_0^2) \Bigr) \ . \label{eqn:EFR}
\end{eqnarray}
One should note that Landau's hypothesis for quasiparticles, Hugenholtz-Van Hove theorem, the virial theorem and conditions of
conserving approximations$^{\cite{LAN}-\cite{UEC}}$ can be exactly satisfied with the effective masses; the effective
masses are naturally defined as the consequence of the requirement of self-consistency.  The structure of
the nonlinear mean-field approximation becomes formally equivalent to that of the linear $\sigma$-$\omega$ Hartree approximation.

\section{Thermodynamic Consistency of the nonlinear mean-field approximation}

The Hugenholtz-Van Hove theorem and equations of motion derived by the theory of conserving approximations are discussed for the proof of
thermodynamic consistency of the current nonlinear approximation.

The energy-momentum tensor operator is given by,
\begin{equation}
{\hat T}^{\mu \nu} = -g^{\mu \nu} {\cal L}_{\scriptscriptstyle MFT}
+ \frac{\partial q_i}{\partial x_{\nu}} \frac{\partial {\cal L}_{\scriptscriptstyle MFT}}
{\partial ( \partial q_i/\partial x^{\mu} )} = {\hat T}_B^{\mu \nu} + {\hat T}_{\sigma}^{\mu \nu} + {\hat T}_{\omega}^{\mu \nu}
+ {\hat T}_{\rho}^{\mu \nu} \ .
\end{equation}
The mean-field expectation values of the baryon, scalar and vector meson operators in the ground state of baryons are defined by,
\begin{equation}
\langle \Psi| {\hat T}_B^{\mu \nu} |\Psi \rangle = \langle \Psi| {\bar \psi}(x) i\gamma^{\mu} \partial^{\nu}
\psi (x) |\Psi \rangle = -i \int\! \frac{d^4 k}{(2\pi)^4} {\rm Tr}(\gamma^{\mu} G(k)) k^{\nu} \ , \label{eqn:TB}
\end{equation}
\begin{eqnarray}
\langle \Psi| {\hat T}_{\sigma}^{\mu \nu} |\Psi \rangle =&&\hspace{-0.6cm} \langle \Psi| \{ -\frac{1}{2}
(\partial_{\lambda} \phi \partial^{\lambda} \phi - m_{\sigma}^2 \phi^2) + \frac{g_{\sigma 3}}{3!} \phi^3 + \frac{g_{\sigma 4}}{4!} \phi^4 \}
g^{\mu \nu} + \partial^{\mu} \phi \partial^{\nu} \phi |\Psi \rangle \nonumber \\
=&&\hspace{-0.6cm}
-i\int\! \frac{d^4 k}{(2\pi)^4} \Bigl\{ (\frac{1}{2} k_{\lambda}^2 - \frac{1}{2} m_{\sigma}^2
-\frac{g_{\sigma 3}}{3!} \phi_0 - \frac{g_{\sigma 4}}{4!}\phi_0^2 ) g^{\mu \nu} - k^{\mu} k^{\nu} \Bigr\} D_{\sigma} (k) \ , \label{eqn:TS}
\end{eqnarray}
\begin{eqnarray}
\langle \Psi| {\hat T}_{\omega}^{\mu \nu} |\Psi \rangle =&&\hspace{-0.6cm} \langle \Psi| \{ \frac{1}{2} (\partial_k
V_{\lambda} \partial^k V^{\lambda} - m_{\omega}^2 V_{\lambda} V^{\lambda}) - \frac{g_{\omega 4}}{4!} (V_{\lambda} V^{\lambda})^2
- \frac{g_{\sigma \omega}}{4} \phi^2 V_{\lambda} V^{\lambda} \} g^{\mu \nu}
- \partial^{\mu} V_{\lambda} \partial^{\nu} V^{\lambda} |\Psi \rangle \nonumber \\
=&&\hspace{-0.6cm} i\int\! \frac{d^4 k}{(2\pi)^4} \Bigl\{ (\frac{1}{2} k_{\lambda}^2 - \frac{1}{2} m_{\omega}^2
- \frac{g_{\omega 4}}{4!} V_0^2 - \frac{g_{\sigma \omega}}{4} \phi_0^2) g^{\mu \nu} - k^{\mu} k^{\nu} \Bigr\} D_{\omega\ a}^{\ \ a} (k) \ , \label{eqn:TV}
\end{eqnarray}
\begin{eqnarray}
\langle \Psi| {\hat T}^{\mu \nu}_{\rho} |\Psi \rangle =&&\hspace{-0.6cm} \langle \Psi| \Bigl\{ \frac{1}{2} (\partial_{\sigma}
{\bf R}_{\lambda} \cdot \partial^{\sigma} {\bf R}^{\lambda} - m_{\rho}^2 {\bf R}_{\lambda} \cdot {\bf R}^{\lambda})
- \frac{g_{\rho 4}}{4!} ({\bf R}_{\lambda} \cdot {\bf R}^{\lambda})^2 -
\frac{g_{\sigma \rho}}{4} \phi^2 ({\bf R}_{\lambda} \cdot {\bf R}^{\lambda}) \nonumber \\
&& -\frac{g_{\omega \rho}}{4} (V_{\lambda} V^{\lambda}) ({\bf R}_{\lambda} \cdot {\bf R}^{\lambda})
\Bigr\} g_{\mu \nu} - \partial_{\mu} {\bf R}^{\lambda} \cdot \partial_{\nu} {\bf R}_{\lambda} |\Psi \rangle \nonumber \\
=&&\hspace{-0.6cm} i\int\! \frac{d^4 k}{(2\pi)^4} \Bigl\{ (\frac{1}{2} k_{\lambda}^2 - \frac{1}{2} m_{\rho}^2
- \frac{g_{\rho 4}}{4!} R_0^2 - \frac{g_{\sigma \rho}}{4} \phi_0^2 - \frac{g_{\omega \rho}}{4} V_0^2) g^{\mu \nu}
- k^{\mu} k^{\nu} \Bigr\} D_{\rho\ a \alpha}^{\ \ a \alpha}(k) \ , \label{eqn:TR}
\end{eqnarray}
where the equations of motion ($\ref{eqn:EQB}$) is used for $\langle \Psi| {\hat T}_B^{\mu \nu} |\Psi \rangle$.  The Roman and Greek indices
in propagators are for spin and isospin degrees of freedom; the repeated indices are summed.  The momentum-space expressions of
the eqs.~($\ref{eqn:TS}$) $\sim$ ($\ref{eqn:TR}$) should be understood as the definition of the nonlinear mean-field
approximation.  The propagators, $D_{\sigma} (k)$, $D_{\omega}^{\ \mu \nu}(k)$ and $D^{\ \mu \nu}_{\rho \ a b} (k)$, are the full
meson-propagators, respectively.  The energy density and pressure of nonlinear $\sigma$, nonlinear $\sigma$-$\omega$-$\rho$ mean-field
approximations$^{\cite{MUL,UEC,BOG}}$ are all reproduced from the eqs.~($\ref{eqn:TB}$) $\sim$ ($\ref{eqn:TR}$) by assuming corresponding interactions.

The energy density, ${\cal E}_{\scriptscriptstyle MFT}$, and pressure, $p_{\scriptscriptstyle MFT}$, can be derived from ($\ref{eqn:TB}$)
to ($\ref{eqn:TR}$), by applying the baryon Green's function $G_{\scriptscriptstyle D} (k)$ and the free nonlinear meson propagators with
effective masses of mesons, $D^{\szero}_{\sigma} (k)$, $D_{\omega}^{{\szero}\ \mu \nu}(k)$ and $D^{{\szero}\ \mu \nu}_{\rho \ a b} (k)$, to tadpole
diagrams$^{\cite{WAL}}$.   One obtains the following results by using the relations: ${\cal E}_{\scriptscriptstyle MFT} = \langle {\hat T}^{00} \rangle$
and $p_{\scriptscriptstyle MFT} = \displaystyle \frac{1}{3} \langle {\hat T}^{ii} \rangle$ ($i =1, 2, 3$, $i$ is summed) as,
\begin{eqnarray}
{\cal E}_{\scriptscriptstyle MFT} =&&\hspace{-0.6cm} \frac{\zeta}{(2\pi)^3} \int^{\kfermi}_0\! d^3k E(k)
+ \frac{m_{\sigma}^2}{2} \phi_0^2 + \frac{g_{\sigma 3}}{3!} \phi_0^3 + \frac{g_{\sigma 4}}{4!} \phi_0^4 
- \frac{m_{\omega}^2}{2} V_0^2 - \frac{g_{\omega 4}}{4!} V_0^4 - \frac{g_{\sigma \omega}}{4} \phi_0^2 V_0^2 \nonumber \\
&&\hspace{-0.6cm} - \frac{(g_{\rho}/2)^2}{m_{\rho}^{\ast 4}} \rho_3^2 \Bigl( \frac{m_{\rho}^2}{2} + \frac{g_{\rho 4}}{4!} R_0^2
+ \frac{g_{\sigma \rho}}{4} \phi_0^2 + \frac{g_{\omega \rho}}{4} V_0^2 \Bigr) \ , \label{eqn:hed}
\end{eqnarray}
\begin{eqnarray}
p_{\scriptscriptstyle MFT} =&&\hspace{-0.6cm} \frac{1}{3}\frac{\zeta}{(2\pi)^3} \int^{\kfermi}_0\! d^3k \frac{k^2}{E^{\ast} (k)}
- \frac{m_{\sigma}^2}{2} \phi_0^2 - \frac{g_{\sigma 3}}{3!} \phi_0^3 - \frac{g_{\sigma 4}}{4!} \phi_0^4
+ \frac{m_{\omega}^2}{2} V_0^2 + \frac{g_{\omega 4}}{4!} V_0^4 + \frac{g_{\sigma \omega}}{4} \phi_0^2 V_0^2 \nonumber \\
&&\hspace{-0.6cm} + \frac{(g_{\rho}/2)^2}{m_{\rho}^{\ast 4}} \rho_3^2 \Bigl( \frac{m_{\rho}^2}{2} + \frac{g_{\rho 4}}{4!} R_0^2
+ \frac{g_{\sigma \rho}}{4} \phi_0^2 + \frac{g_{\omega \rho}}{4} V_0^2 \Bigr) \ , \ \label{eqn:hpre}
\end{eqnarray}
and note that mean fields of mesons are also related with baryon density as, $\phi_0 = (M - \Mstar (\kfermi))/g_{\sigma}$,
$V_0 = (g_{\omega}^{\ast}/m_{\omega}^{\ast 2}) \rhoB$ and $R_0 = (g_{\rho}^{\ast}/2m_{\rho}^{\ast 2}) \rho_3$.  The
vacuum expectation value of the energy-momentum tensor must be subtracted when the finite physical energy densities
are extracted from eqs.~($\ref{eqn:TS}$) $\sim$ ($\ref{eqn:TR}$)$^{\cite{WAL}}$.

Now, thermodynamic consistency of the current approximation can be shown explicitly.  By adding ($\ref{eqn:hed}$) and ($\ref{eqn:hpre}$), one
can directly calculate that the energy density, pressure and single particle energy, (${\cal E}_{\scriptscriptstyle MFT}$,
$p_{\scriptscriptstyle MFT}$, $E(k)$), satisfy the thermodynamic relation:
\begin{equation}
{\cal E}_{\scriptscriptstyle MFT} + p_{\scriptscriptstyle MFT} = \rhoB E(\kfermi) \ . \label{eqn:THR}
\end{equation}
In order to obtain ($\ref{eqn:THR}$), the single particle energy, $E(k) = E^{\ast} (k) - \Sigma^0 (\kfermi)$, is used to partially integrate
the first term of ($\ref{eqn:hed}$).  By adding the result to the first term of ($\ref{eqn:hpre}$), one obtains the right-hand side of
($\ref{eqn:THR}$); the rest of the terms cancel one another.   The equality of chemical potential and the single particle energy, $\mu = E(\kfermi)$,
is also proved by differentiating the energy density, ${\cal E}_{\scriptscriptstyle MFT}$, with respect to the baryon density, $\rhoB$, and
using equations of motion for mesons~($\ref{eqn:EMM}$); it results in,
\begin{equation}
\mu = \frac{\partial {\cal E}_{\scriptscriptstyle MFT}}{\partial \rhoB} = E(\kfermi) \ . \label{eqn:HVT}
\end{equation}
The result ($\ref{eqn:HVT}$) is essential for the proof of the Hugenholtz-Van Hove theorem to an approximation and for the definition of the density of
nuclear matter saturation, where the pressure vanishes: $p_{\scriptscriptstyle MFT} = 0$.

The self-consistency of effective masses of mesons, $m_{\sigma}^{\ast}$, $m_{\omega}^{\ast}$ and $m_{\rho}^{\ast}$, is again proved by
the condition of conserving approximations$^{\cite{BAY,UEC}}$.  The functional derivative of energy density,
${\cal E}_{\scriptscriptstyle MFT}(\phi_0, V_0, R_0, n_i)$, with respect to the baryon number distribution, $n_i$, is given by:
\begin{equation}
\frac{\delta {\cal E}_{\scriptscriptstyle MFT}}{\delta n_i} = E(k_i) +
\sum_i \Bigl( \frac{\delta {\cal E}_{\scriptscriptstyle MFT}}{\delta \phi_0} \frac{\delta \phi_0}{\delta n_i} 
+ \frac{\delta {\cal E}_{\scriptscriptstyle MFT}}{\delta V_0}
\frac{\delta V_0}{\delta n_i} + \frac{\delta {\cal E}_{\scriptscriptstyle MFT}}{\delta R_0} \frac{\delta R_0}{\delta n_i} \Bigr) \ . \label{eqn:dedn}
\end{equation}
The self-consistent condition of the theory of conserving approximations requires: $\displaystyle
\frac{\delta {\cal E}_{\scriptscriptstyle MFT}}{\delta \phi_0} = 0$, $\displaystyle \frac{\delta {\cal E}_{\scriptscriptstyle MFT}}{\delta V_0} = 0$
and $\displaystyle \frac{\delta {\cal E}_{\scriptscriptstyle MFT}}{\delta R_0} = 0$$^{\cite{UEC}}$.  The condition independently generates
meson equations of motion and determines self-energies of the approximation.  The self-energies calculated by propagators,
$(\ref{eqn:SS}) \sim (\ref{eqn:0R})$, and by the condition of conserving approximations become equivalent, only if the effective masses of
mesons are given by~($\ref{eqn:EFR}$).  The nonlinear mean-field approximation becomes thermodynamically consistent, relativistic,
field-theoretical approximation, with the effective masses of mesons and coupling constants.  The current functional differential approach can be
applied to Hartree-Fock, Ring$^{\cite{UEC}}$ and other approximations, in order to obtain exact solutions which maintain the HV theorem and
Landau's requirement of quasiparticles.

\section{Properties of nuclear matter and neutron stars}

The linear $\sigma$-$\omega$ Hartree approximation (LHA)$^{\cite{WAL}}$, and the current nonlinear Hartree approximations,
denoted as NHA$^{2.50}$ and NHA$^{2.00}$ which generate $M_{max} = 2.50\ M_{\odot}$ and $2.00\ M_{\odot}$, will be compared
in this section.  They are evaluated numerically by solving equations of motion iteratively with a given density $\rhoB$; then, the single
particle energy, energy density and pressure are calculated.  Since the Hugenholtz-Van Hove theorem is maintained,
Fermi-liquid properties of nuclear matter are unambiguously determined, and properties of neutron stars are evaluated by employing
the equation of state for neutron matter.  One should note that self-consistency defined by the HV theorem, the virial theorem and
conditions of conserving approximations are essential to calculate properties of infinite nuclear and neutron matter.

The coupling constants are adjusted in order to produce saturation properties, ${\cal E}/\rhoB - M = -15.75$ MeV, at $\kfermi = 1.30$ fm$^{-1}$,
the symmetry energy $a_4 = 35.0$ MeV, and simultaneously, the minimum value of compressibility, $K$, is investigated to the maximum mass,
$M_{max} = 2.50\ M_{\odot}$ and $2.00\ M_{\odot}$, of neutron stars$^{\cite{AKM,LAT}}$.  The calculation is repeated until required values are
obtained.  Contrary to an expectation, the adjustment of nonlinear coupling constants are not free at all because they are restricted by
the coupled equations of motion and constraints of thermodynamic consistency; the values of nonlinear coupling constants are confined within a small
region.  Consequently, the equation of state and properties of nuclear matter compatible with the mass of hadronic neutron stars will be explicitly
constrained by the density-dependent effective masses and coupling constants.  The results should be analyzed quantitatively by
nuclear experimental data in order to examine the validity of nonlinear mean-field models.

The coupling constants which increase or decrease the values of both $K$ and $a_4$ will have the same tendency to the value of $M_{max}$, and it
depends on respective coupling constants how much they affect physical quantities.  The $\sigma$-$\omega$ mixing interaction term,
$g_{\sigma \omega}$, is an exception that decreases the value of K but increases the values of both $a_4$ and $M_{max}$, which indicates the
$\sigma$-$\omega$ mixing interaction may simulate distinct many-body interactions at normal density.  Even if one may assume the value of $g_{\sigma 3}$
as $10^2 \sim 10^3$, the $g_{\sigma 3}$-term produce small contributions to physical quantities which are readily recovered by changing
$g_{\sigma 4}$ and $g_{\omega 4}$.  The properties of nuclear matter and neutron stars are modified mainly by changing $g_{\sigma 4}$,
$g_{\omega 4}$ and the mixing term $g_{\sigma \omega}$.  The coupling constant, $g_{\sigma \rho N}$, contributes mainly to $a_4$ and $M_{max}$.

The compressibility is calculated by
\begin{equation}
K = 9\rhoB \frac{\partial^2 {\cal E}}{\partial \rhoB^2} = 9\rhoB \biggl( \frac{\partial \mu}{\partial \rhoB} \biggr) \label{eqn:K} \ ,
\end{equation}
where $\mu$ is the chemical potential and equal to the Fermi energy, $\mu = E(\kfermi)$, since HV theorem and Landau's hypothesis for
quasiparticles are maintained exactly.  The binding energy curves given by the linear $\sigma$-$\omega$ Hartree
approximation (LHA)$^{\cite{WAL}}$ and the current nonlinear Hartree approximations, NHA$^{2.50}$ and NHA$^{2.00}$, corresponding to
the maximum masses of neutron stars $M_{max} = 2.50\ M_{\odot}$ and $M_{max} = 2.00\ M_{\odot}$, are compared in the Fig.~2.

Coupling constants and Fermi-liquid properties of nuclear matter and neutron stars are listed in the Table~I.  The linear Hartree
approximation (LHA) gives $K=530$ MeV and $M_{max} = 3.09\ M_{\odot}$.  The nonlinear Hartree approximation,
NHA$^{2.50}$, shows that the lower bound of compressibility is about $K \sim 340 $ MeV, whereas NHA$^{2.00}$ exhibits that the lower bound
of compressibility is about $K \sim 180 $ MeV.  The results are quite distinctive that it should be investigated by studying experimental data of
nuclear and neutron matter.

The effective mass of nucleons, $\Mstar/M$, and the effective masses of mesons, $m_{\sigma}^{\ast}/m_{\sigma}$ and
$m_{\omega}^{\ast}/m_{\omega}$, of NHA$^{2.50}$ and NHA$^{2.00}$ in nuclear matter are shown in the Figs.~3a and~3b by comparing with the linear
$\sigma$-$\omega$ Hartree approximation (LHA).  The effective masses of nucleons and mesons are slightly increased at saturation;
the results of NHA$^{2.50}$ are $\Mstar/M = 0.62$, $m_{\sigma}^{\ast}/m_{\sigma} = 1.06$ and $m_{\omega}^{\ast}/m_{\omega} = 1.05$ for $K=340$ MeV,
$a_4 = 35.0$ MeV and $M_{max} = 2.50\ M_{\odot}$.  The NHA$^{2.00}$ generates: $\Mstar/M = 0.72$, $m_{\sigma}^{\ast}/m_{\sigma} = 1.20$
and $m_{\omega}^{\ast}/m_{\omega} = 1.17$ for $K=186$ MeV, $a_4 = 35.0$ MeV and $M_{max} = 2.00\ M_{\odot}$.  The effective masses of neutron matter
for NHA$^{2.50}$ and NHA$^{2.00}$ are shown in the Figs.~3c and~3d.

Notice that the effective mass of nucleons is given by $\Mstar = M - g_{\sigma} \phi_0$ and the sigma mean-field, $\phi_0$, increases rapidly
in the LHA, resulting in the rapid decrease of $\Mstar$, but the decrease of $\Mstar$ is smoothed and counterbalanced by the increase of meson
masses in the nonlinear approximation.  The density-dependences of effective constants, $g^{\ast}_{\omega}$ and $g^{\ast}_{\rho}$, are shown
in the Fig.~4.  Similar to the increase of meson masses, the nonlinear interactions increase the values of effective coupling constants. However,
nonlinear interactions mainly contribute to physical quantities through effective masses of baryons and mesons.  The $\sigma\omega N$ effective
coupling constant changes only less than $3\%$: $g_{\omega}^{\ast}/g_{\omega} = 1.01 \sim 1.03$ at saturation density, when the empirical data
are altered from NHA$^{2.50}$ to NHA$^{2.00}$.  This shows that coupling constants are insensitive to nonlinear interactions in order to determine
experimental values in the level of Hartree approximation.   The $\sigma\rho N$ effective coupling constant changes $g_{\rho}^{\ast}/g_{\rho} = 1.07
\sim 1.14$ at saturation density.  The large values of $g_{\rho}^{\ast}/g_{\rho}$ are needed in order to reproduce the value of symmetry energy,
$a_4=35.0$ MeV, in the Hartree approximation.  However, the important contributions to the symmetry energy also come from the exchange
interactions$^{\cite{MAT}}$, which will compensate the value of the symmetry energy and may lead to the negligible value of
$g_{\rho}^{\ast}/g_{\rho}$ at saturation density.

The nonlinear interactions constrain effective masses of baryons and mesons, $\Mstar$, $m_{\sigma}^{\ast}$, $m_{\omega}^{\ast}$, $m_{\rho}^{\ast}$
and effective coupling constants, $g^{\ast}_{\omega}$, $g^{\ast}_{\rho}$.  Especially, the results should be compared with the rest masses of
hadrons ($M$, $m_{\omega}$, $m_{\rho}$, $m_{\pi}$, $\cdots$) in order to check nuclear experimental data whether masses of mesons are modified
inside of nucleus or not.  The quantitative study of effective masses by nuclear experimental data is essential to conclude the validity of
nonlinear mean-field models.

The symmetry energy is calculated by
\begin{equation}
a_4 = \frac{1}{2}\rhoB \biggl[ \biggl[ \frac{\partial^2 {\cal E}}{\partial \rho_3^2} \biggr]_{\rhoB} \biggr]_{\rho_3 = 0} \label{eqn:RA4} \ ,
\end{equation}
where $\rho_3$ is the difference between the proton and neutron density: $\rho_3 = \rho_p - \rho_n = (k^3_{F_p} - k^3_{F_n})/3\pi^2$ at
a fixed baryon density, $\rhoB = \rho_p + \rho_n = 2 \kfermi^3/3\pi^2 $.  The $\rho$-meson mean-field $R_0$ and self-energies are calculated by,
\begin{equation}
\frac{g_{\rho}^{\ast}}{2} R_0 = - \Sigma_{\rho p}^0 = \Sigma_{\rho n}^0 = \frac{g_{\rho}^{\ast 2}}{4m_{\rho}^{\ast 2}} \rho_3 \label{eqn:SPN} \ ,
\end{equation}
where $g_{\rho}^{\ast} = g_{\rho} + 2 g_{\sigma \rho N} \phi_0$.  The effective mass of $\rho$-meson is given by
\begin{equation}
m_{\rho}^{\ast 2} = m_{\rho}^2(1 + \frac{g_{\rho 4}}{6m_{\rho}^2} R_0^2 + \frac{g_{\sigma \rho}}{2m_{\rho}^2} \phi_0^2
+ \frac{g_{\omega \rho}}{2m_{\rho}^2} V_0^2) \ , \label{eqn:SMR}
\end{equation}
and it is self-consistently determined by the eq.~($\ref{eqn:SPN}$).  In symmetric nuclear matter ($\rho_3 = 0$), the $\rho$-meson mean-field,
$R_0$, $\sigma$-$\rho$ and $\omega$-$\rho$ interactions vanish, and $m_{\rho}^{\ast} = m_{\rho}$ will be restored.  The effective mass of
$\rho$-meson is similar to the bare mass: $m_{\rho}^{\ast} \sim m_{\rho}$ at saturation density, since contributions to the effective mass of
$\rho$-meson from nonlinear interactions are negligible in the level of mean-field approximation.  Although the coupling constant, $g_{\rho_4}$,
is fixed from the beginning according to the ref.~${\cite{MUL}}$, one can see from the eq.~($\ref{eqn:SMR}$) that the nonlinear self-interaction
of $\rho$-meson will not contribute to physical qunatities at all, since the contribution to $m_{\rho}^{\ast}$ from the $g_{\rho 4}$-term in
($\ref{eqn:SMR}$) is negligible.  The symmetry energy, $a_4$, versus Fermi-momentum, $\kfermi$, is shown in the Fig.~5.  Nonlinear
interactions restricted by self-consistency and properties of nuclear and neutron matter do not quite change the value of symmetry energy.

It is difficult to reproduce required values of $M_{max}$ and $a_4$ in the numerical calculation, since properties at nuclear matter saturation,
coupled equations of motion and self-consistency will confine adjustable parameters strictly.  Coupling constants should be carefully adjusted
so that one can obtain solutions in the density range, $0 < \kfermi \le 3.0$ fm$^{-1}$, where the equation of
state is important to determine the mass of neutron stars.  The conditions of conserving approximation and consequently, self-consistent condition,
$\partial {\cal E}/\partial n_i = E (k_i)$, are essential to evaluate eqs.~($\ref{eqn:K}$) $\sim$ ($\ref{eqn:RA4}$) of compressibility and symmetry energy.

The equations of state for LHA, NHA$^{2.50}$ and NHA$^{2.00}$ of neutron matter are shown in the Fig.~6.  The masses of neutron stars are calculated
by Tollman-Oppenheimer-Volkoff (TOV) equation$^{\cite{WEI,HEI}}$ and shown in the Fig.~7, as a function of a central energy density,
${\cal E}_c$.  The data of NHA$^{2.00}$ indicate that the equation of state which produces compressibility, $K > 180$ MeV, is necessary in order to
reproduce all the observed masses of hadronic neutron stars; $K \sim 180$ MeV is the lower bound to support all the hadronic neutron stars
in the nonlinear $\sigma$-$\omega$-$\rho$ mean-field approximation.

The moment of inertia, $I$, of neutron stars is calculated by$^{\cite{HEI,HAR}}$,
\begin{equation}
I = \frac{8\pi}{3} \int^{R}_0 \! dr \frac{r^4 ({\cal E}(r) + p(r))\ e^{-\nu (r)/2}}{( 1 - 2 G M(r)/r)^{1/2}} \label{eqn:IMS} \ ,
\end{equation}
where $G$ is the gravitational constant.  The physical constants are set as $\hbar = c = M_{\odot} = 1$, and so, $G = G M_{\odot}/c^2 = 1.476$ km;
the moment of inertia is given by the unit of $M_{\odot}\cdot$km$^2$.  The metric function, $\nu (r)$, is given by
\begin{equation}
\frac{d \nu (r)}{dr} = \frac{2 G M(r) + 4\pi r^3 G p(r)}{ r (r - 2 G M(r) )} \ ,
\end{equation}
with $d\nu (r)/dr \rightarrow 0 \ (r \rightarrow 0)$, and the boundary condition is
\begin{equation}
\nu (R) = \log ( 1 - 2 G M(R)/R ) \ ,
\end{equation}
at the surface of a neutron star, $r = R$.  The masses of neutron stars and moments of inertia are shown in Fig.~8, Fig.~9 and Fig.~10.  The
approximation, NHA$^{2.50}$, gives, $K = 340$ MeV and $a_4 = 35.0$ MeV, the neutron star mass, $M_{max} = 2.50\ M_{\odot}$, the moment of inertia,
$I = 368\ M_{\odot}{\rm km}^2$, and the radius, $R = 13.2\ {\rm km}$; that is, the lower bound of compressibility is expected to be $K \sim 340$ MeV
($M_{max} > 2.50\ M_{\odot}$).  The approximation, NHA$^{2.00}$, generates $K= 186$ MeV and $a_4 = 35.0$ MeV, the neutron star mass,
$M_{max} = 2.00\ M_{\odot}$, $I = 192\ M_{\odot}{\rm km}^2$ and $R = 11.9\ {\rm km}$.

The lower bound of compressibility changes about 50\% ($K = 340 \rightarrow 170$ MeV), as the maximum mass of neutron stars is changed
20\% ($M_{max} = 2.50 \rightarrow 2.00\ M_{\odot}$).  Compared $K$ with $a_4$, the compressibility changes about 20\%
($K = 340 \rightarrow 270$ MeV), as the symmetry energy, $a_4$, changes about 8\% ($35.0 \rightarrow 32.5$ MeV).  The properties of nuclear
and neutron matter are closely related to each other; the accumulation of data and accurate measurement of experimental values are very important
to understand many-body systems of hadrons at nuclear density, as well as systems at high-energy and high-density.

All the results are consistent with the analysis of acceptable configuration of the central energy and maximum masses of neutron
stars$^{\cite{LAT}}$.  The approximations LHA, NHA$^{2.50}$, and NHA$^{2.00}$ produce distinct results for properties of nuclear and neutron
stars, which may be distinguished in the course of the accurate examinations of nuclear experimental data and accumulating data of neutron
stars.  Since properties of nuclear matter and neutron stars are closely related, the theoretical analysis of self-consistency to an approximation
is essential to calculate physical quantities.  One should note that retardation corrections may break self-consistency to an approximate
calculations and cause serious discrepancy in self-consistent calculations$^{\cite{UEC}}$.  The current technique of constructing a conserving
approximation gives a method to extend an approximation self-consistently.

\section{Conclusions and Remarks}

We have employed a nonlinear mean-field model in order to simulate complicated many-body interactions of hadrons and studied Fermi-liquid
properties of nuclear matter and neutron stars.  Nonlinear interactions renormalized self-consistently with thermodynamic consistency
result in the introduction of effective masses and effective coupling constants of nucleons and mesons.  The contributions of nonlinear
interactions to physical quantities (effective masses, $M^{\ast}$, $m_{\sigma}^{\ast}$, $m_{\omega}^{\ast}$, $m_{\rho}^{\ast}$, compressibility,
$K$ and symmetry energy, $a_4$) and neutron stars (maximum mass $M_{max}$, central energy density ${\cal E}_c$, moment of inertia $I$,
and radius $R$) are discussed and mutual relations of the physical quantities are examined.  The nonlinear mean-field models seem to simulate properties
of nuclear and neutron matter reasonably well$^{\cite{MUL}}$.  However, since the effective masses of mesons and nucleons are self-consistently
confined in order to maintain HV theorem, thermodynamic consistency$^{\cite{BAY,UEC}}$, properties of infinite matter and coupled
equations of motion for mesons, the values of nonlinear coupling constants are not free parameters to adjust at all.

The nonlinear interactions are constrained with self-consistency and generate rather restricted values of properties of nuclear and neutron
matter.  The softer equation of state will decrease the value of compressibility, $K$, and simultaneously, decrease the maximum mass of neutron
stars, $M_{max}$ and symmetry energy, $a_4$.  Nonlinear interactions will connect physical quantities closely to one another that
it is difficult to reproduce experimentally desired values of $K$, $a_4$ and $M_{max}$.  A fine adjustment of nonlinear coupling constants
is needed to produce the experimentally required values of $K$, $a_4$ and $M_{max}$.  Therefore, it is important to examine quantitatively
by nuclear experimental data, whether density-dependent effective masses ($\Mstar$, $m^{\ast}_{\omega}$,
$m^{\ast}_{\pi}$, $m^{\ast}_{\rho}$) and coupling constants exist or not.  The results will clarify the
validity of nonlinear interactions and the relation between Fermi-liquid properties of nuclear and neutron stars.

The nonlinear self-interactions of $g_{\sigma 4}$ and $g_{\omega 4}$, mixing-interaction of $g_{\sigma \omega}$ terms are especially important to
determine the binding energy, compressibility and the mass of neutron stars, but the nonlinear
interactions yield little contribution to symmetry energy in the level of Hartree approximation.  The density-dependence of effective coupling
constants, $g_{\omega}^{\ast}$ and $g_{\rho}^{\ast}$, are not so important to reproduce the symmetry energy $a_4$ and $M_{max}$ of neutron
stars.  The existence of solutions to coupled equations of motion for mesons and convergences of properties of nuclear matter in the density range,
$\kfermi = 1.0 \sim 3.0$ fm$^{-1}$, should be carefully checked in order to calculate properties of neutron stars.  The values of nonlinear
coupling constants might be studied as renormalized values of many-body interactions, such as Hartree-Fock, Ring, Ladder and other effective
approximations with vertex interactions; the effective masses and coupling constants should be investigated further in terms of many-body interactions.

A mixed-phase of quark-hadron star might be expected.  A hadron-quark matter mixed-phase equation of state which is defined by MIT-bag model,
for example, could produce the maximum observed mass: $M_{max} = 2.50\ M_{\odot}$$^{\cite{UEC3}}$.  However, even if the experimental value
of compressibility and the maximum mass of neutron stars are expected to be $K \sim 200$ MeV$^{\cite{YOU}}$ and $M_{max} \sim 2.00\ M_{\odot}$,
the nonlinear $\sigma$-$\omega$-$\rho$ mean-field approximation indicates that it is possible to generate neutron stars which might be pure
hadron-stars.  In order to distinguish the inner structure of neutron stars from the observed experimental data, it is important to
analyze effective masses of baryons and mesons from nuclear experimental data as precisely as possible.

The current approach by employing the theory of conserving approximations is essential to check self-consistency and calculate equations of state,
Fermi-liquid properties and Landau parameters$^{\cite{MAT}}$ of nuclear matter.  It provides us with a method to construct thermodynamically consistent
approximations.  The nonlinear mean-field approximation suggests that nonlinear interactions of mesons and baryons be perceived as {\it effective masses}
and {\it effective coupling constants}.  These renormalized constants are essential and naturally required for self-consistency to an
approximation and might be used to examine the validity of nonlinear models of hadrons.  The accumulation of data and the analysis of neutron stars
should be investigated further in order to elucidate the validity of nonlinear mean-field models and theory of nuclear models.

\newpage

\centerline{ \bf Table I.  Properties of nuclear matter and neutron stars }

The results of the linear $\sigma$-$\omega$ Hartree approximation (LHA)$^{\cite{WAL}}$, and the current nonlinear
Hartree approximations, NHA$^{2.50}$ and NHA$^{2.00}$ for $M_{max} = 2.50\ M_{\odot}$ and $2.00\ M_{\odot}$, are shown respectively.  $\Mstar$ is
the effective mass of nucleon; $m_{\sigma}^{\ast}$, $m_{\omega}^{\ast}$ and $m_{\rho}^{\ast}$ are effective masses of $\sigma$, $\omega$ and
$\rho$ mesons, respectively.  $K$ is compressibility and $a_4$ is symmetry energy.   The masses are fixed as
$M = 939$ MeV, $m_{\sigma} = 550$ MeV, $m_{\omega} = 783$ MeV and $m_{\rho} = 770$ MeV$^{\cite{WAL}}$.  $M_{max}$ is the maximum mass
in the solar mass unit ($M_{\odot}$) and ${\cal E}_C$ ($10^{15}\ {\rm g/cm^3})$ is the central energy density; $I$ is the inertial mass
($M_{\odot}\cdot km^2$) and $R$ ($km$) is the radius of a neutron star.  The coupling constants are adjusted to reproduce the values:
${\cal E}/\rhoB - M = -15.75$ MeV, at $\kfermi = 1.30$ fm$^{-1}$, and $a_4 = 35.0$ MeV; the maximum masses of neutron stars,
$M_{max} = 2.50$ $M_{\odot}$ and $M_{max} = 2.00$ $M_{\odot}$$^{\cite{WAL}-\cite{LAT}}$.  The values of vertex coupling constants are,
$g_{\sigma\omega N} = 0.0050$ and $g_{\sigma\rho N} = 0.0080$ for NHA$^{2.50}$; $g_{\sigma\omega N} = 0.0120$ and $g_{\sigma\rho N} = 0.0215$
for NHA$^{2.00}$.  The coupling constants, $g_{\rho}$ and $g_{\rho 4}$, are fixed$^{\cite{WAL}-\cite{MUL}}$.

\begin{center}
\arrayrulewidth=1.0pt
\doublerulesep=0pt 
\begin{tabular}{lcccccccccl} \\ \hline\hline
          &  $g_{\sigma}$   &  $g_{\omega}$  &  $g_{\rho}$ & $g_{\sigma 3}$ ({\small MeV}) &\ $g_{\sigma 4}$ \ & 
$g_{\omega 4}$ & \ $g_{\rho 4}$ & $g_{\sigma \omega}$ & $g_{\sigma \rho}$ & \ \ $g_{\omega \rho}$ \\ \hline\hline
LHA           & 11.071 & 13.795 & 8.077  & 0.0 & 0.0 & 0.0 & 0.00  & 0.0 & 0.0  &  0.0 \\
NHA$^{2.50}$  & 10.647 & 12.863 & 8.077  & 100 & 280 & 280 & 4.00  & 70  & 2.0  &  2.0 \\
NHA$^{2.00}$  & 10.263 & 11.737 & 8.077  & 480 & 1700 & 1700 & 4.00 & 440 & 21.0 & 21.0 \\
\hline\hline
\end{tabular}
\end{center}

\begin{center}
\arrayrulewidth=1.0pt
\doublerulesep=0pt
\begin{tabular}{lccccccccc} \\ \hline\hline
          &  $M^{\ast}/M$   &  $m_{\sigma}^{\ast}/m_{\sigma}$  & $m_{\omega}^{\ast}/m_{\omega}$
          &  $K$ (MeV) & $a_4$ (MeV) &$M_{max}$ & ${\cal E}_c$ &  $ I $ \quad & R (km) \\ \hline\hline
LHA           & 0.54 & 1.00 & 1.00 & 530 & 35.0 & 3.09 & 1.25 & 717 & 14.5 \\
NHA$^{2.50}$  & 0.62 & 1.06 & 1.05 & 340 & 35.0 & 2.50 & 1.53 & 368 & 13.2 \\
NHA$^{2.00}$  & 0.72 & 1.20 & 1.17 & 186 & 35.0 & 2.00 & 1.94 & 192 & 11.9 \\
\hline\hline
\end{tabular}
\end{center}

\newpage

\centerline{\Large{\bf Figure Captions}}

\begin{description}

\item[Fig.~1a.] The scalar self-energy, $\Sigma^s$, is given by renormalized effective masses and nonlinear coupling terms.  The vertices
are $-ig_{\sigma\omega N} V_0 \gamma^0$ for $\sigma\omega N$-vertex and $-ig_{\sigma\rho N} R_0 \tau_3 \gamma^0$ for $\sigma\rho N$-vertex.

\item[Fig.~1b.] The omega and rho meson sources to generate $\Sigma^0$.  The vertex factors are $-ig_{\omega}^{\ast} \gamma^{\mu}$ for
$\omega N$ and $-i(g_{\rho}^{\ast}/2) \tau_3 \gamma^{\mu}$ for $\rho N$ couplings.

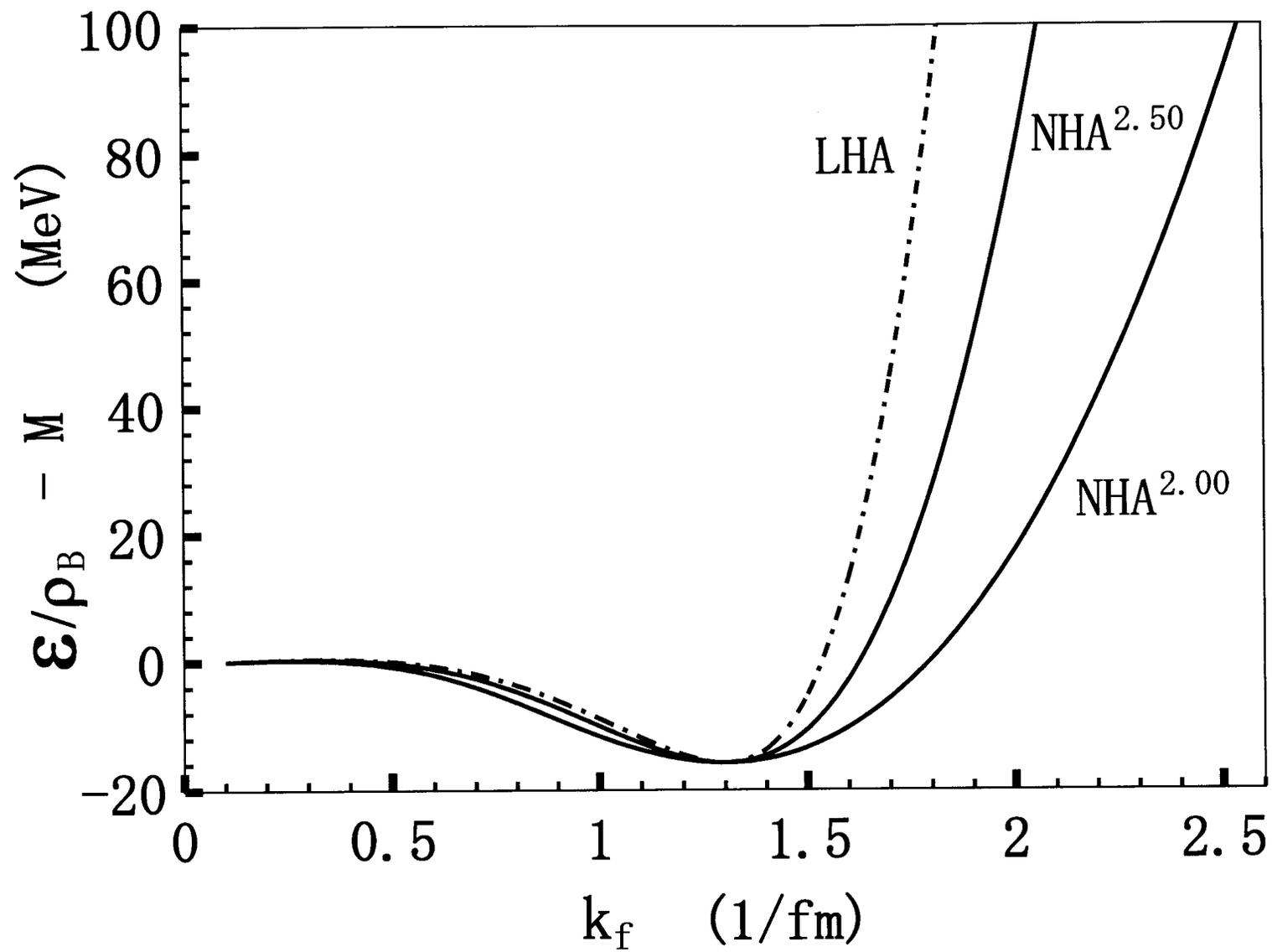
\item[Fig.~2.] The binding energy curves of symmetric nuclear matter.  The linear $\sigma$-$\omega$ Hartree approximation (LHA), and the
nonlinear Hartree approximations, NHA$^{2.50}$ and NHA$^{2.00}$ which produce the maximum masses of neutron stars:
$M_{max} = 2.50\ M_{\odot}$ and $M_{max} = 2.00\ M_{\odot}$, are shown respectively (see, the table~I).

\item[Fig.~3a.] The effective nucleon mass $\Mstar/M$ (solid-line), the effective scalar meson mass, $m_s^{\ast}/m_s$ (dotted-line), and the
effective vector meson mass, $m_{\omega}^{\ast}/m_{\omega}$ (dot-dashed line), for the binding energy of nuclear matter, NHA$^{2.50}$
($K = 340$ MeV, $M_{max} = 2.50\ M_{\odot}$).

\item[Fig.~3b.] The effective masses of nucleons and mesons for nuclear matter, NHA$^{2.00}$ ($K = 186$ MeV, $M_{max} = 2.00\ M_{\odot}$).

\item[Fig.~3c.] The effective masses of nucleons and mesons for neutron matter, NHA$^{2.50}$ ($K = 340$ MeV, $M_{max} = 2.50\ M_{\odot}$).  The
degeneracy factor is $\zeta = 2$, with the same coupling constants as those of Fig.~3a (see, the Table~I).

\item[Fig.~3d.] The effective masses of nucleons and mesons for neutron matter, NHA$^{2.00}$ ($K = 186$ MeV, $M_{max} = 2.00\ M_{\odot}$).  The
degeneracy factor is $\zeta = 2$, with the same coupling constants as those of Fig.~3b (see, the Table~I).

\item[Fig.~4.] The effective coupling constants, $g_{\omega}^{\ast}/g_{\omega}$ (solid line) and $g_{\rho}^{\ast}/g_{\rho}$ (dashed line).  In
the solid and dashed lines, the lower lines are for NHA$^{2.50}$ and the upper lines for NHA$^{2.00}$.

\item[Fig.~5.] The symmetry energies, LHA, NHA$^{2.50}$ and NHA$^{2.00}$, versus Fermi-momentum, $\kfermi$,
for Table~I are shown.

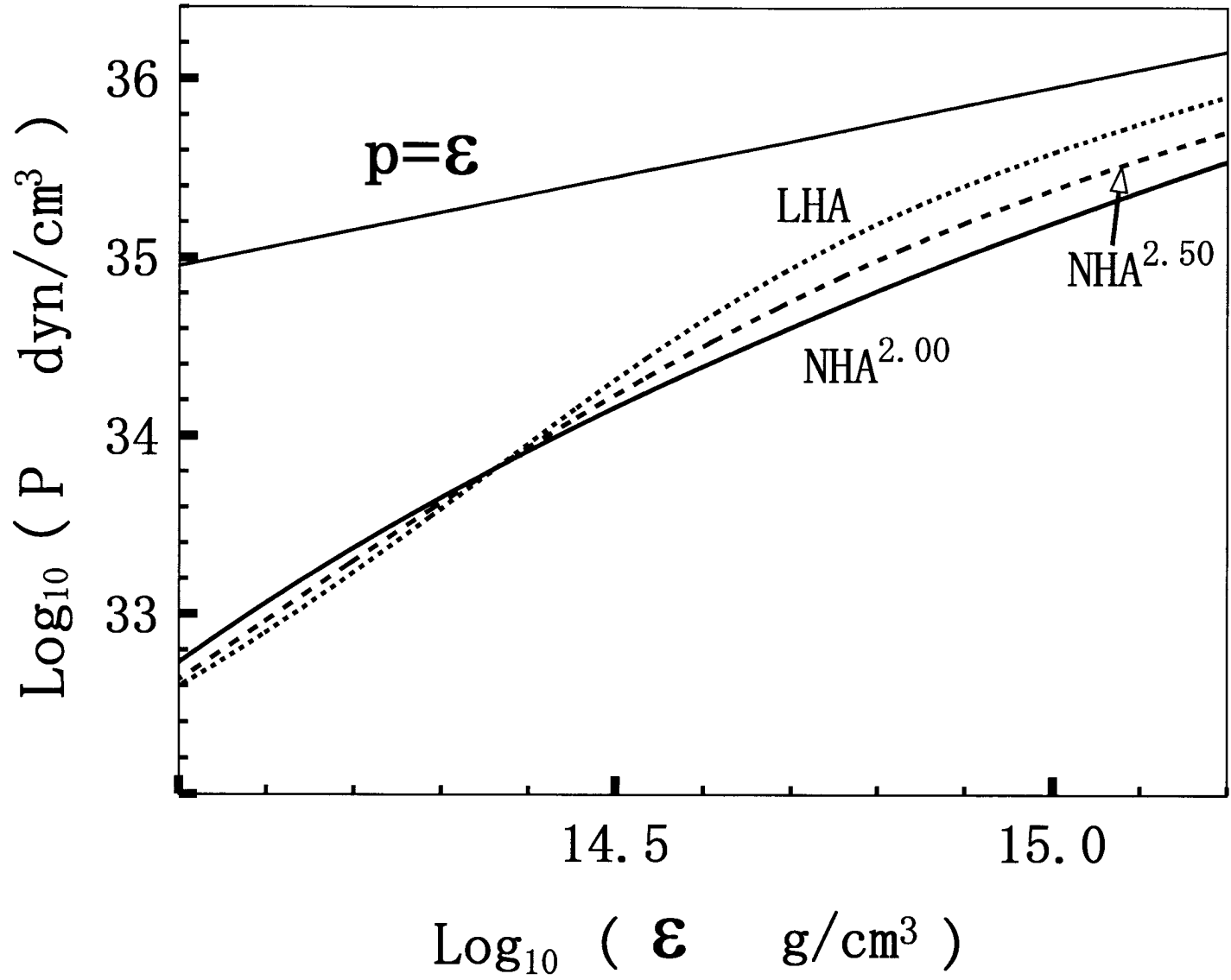
\item[Fig.~6.] The equations of state for neutron matter.  $p = {\cal E}$ is the causal limit of the equation of state.

\item[Fig.~7.] Neutron star masses as functions of the central energy density, ${\cal E}_c$.  The lines are LHA ($M_{max} = 3.09\ M_{\odot}$,
${\cal E}_c = 1.25 \times 10^{15}$ g/cm$^3$), NHA$^{2.50}$ ($M_{max} = 2.50\ M_{\odot}$, ${\cal E}_c = 1.53 \times 10^{15}$ g/cm$^3$), and
NHA$^{2.00}$ ($M_{max} = 2.00\ M_{\odot}$, ${\cal E}_c = 1.94 \times 10^{15}$ g/cm$^3$), respectively.

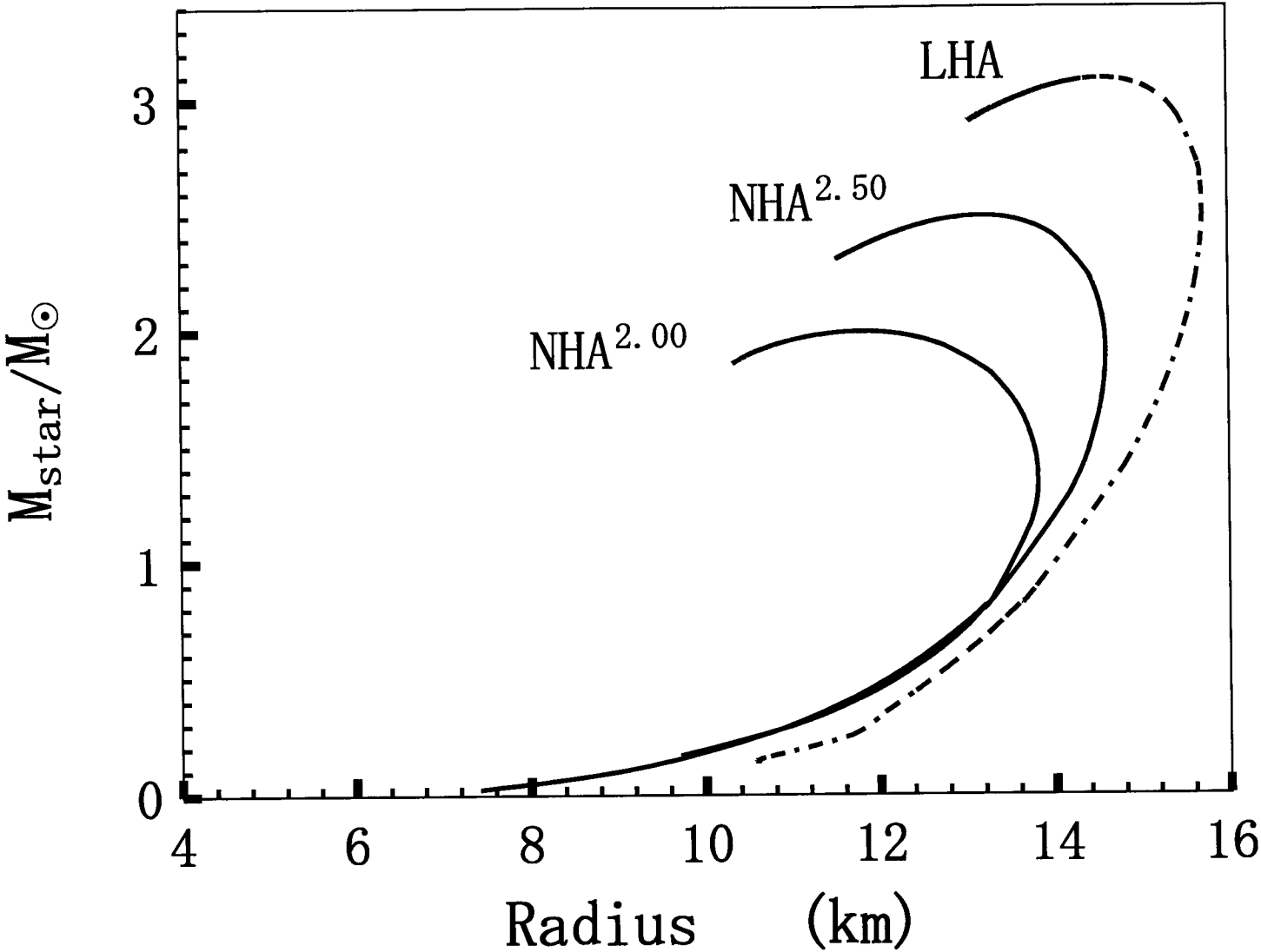
\item[Fig.~8.] The masses and radii of neutron stars.

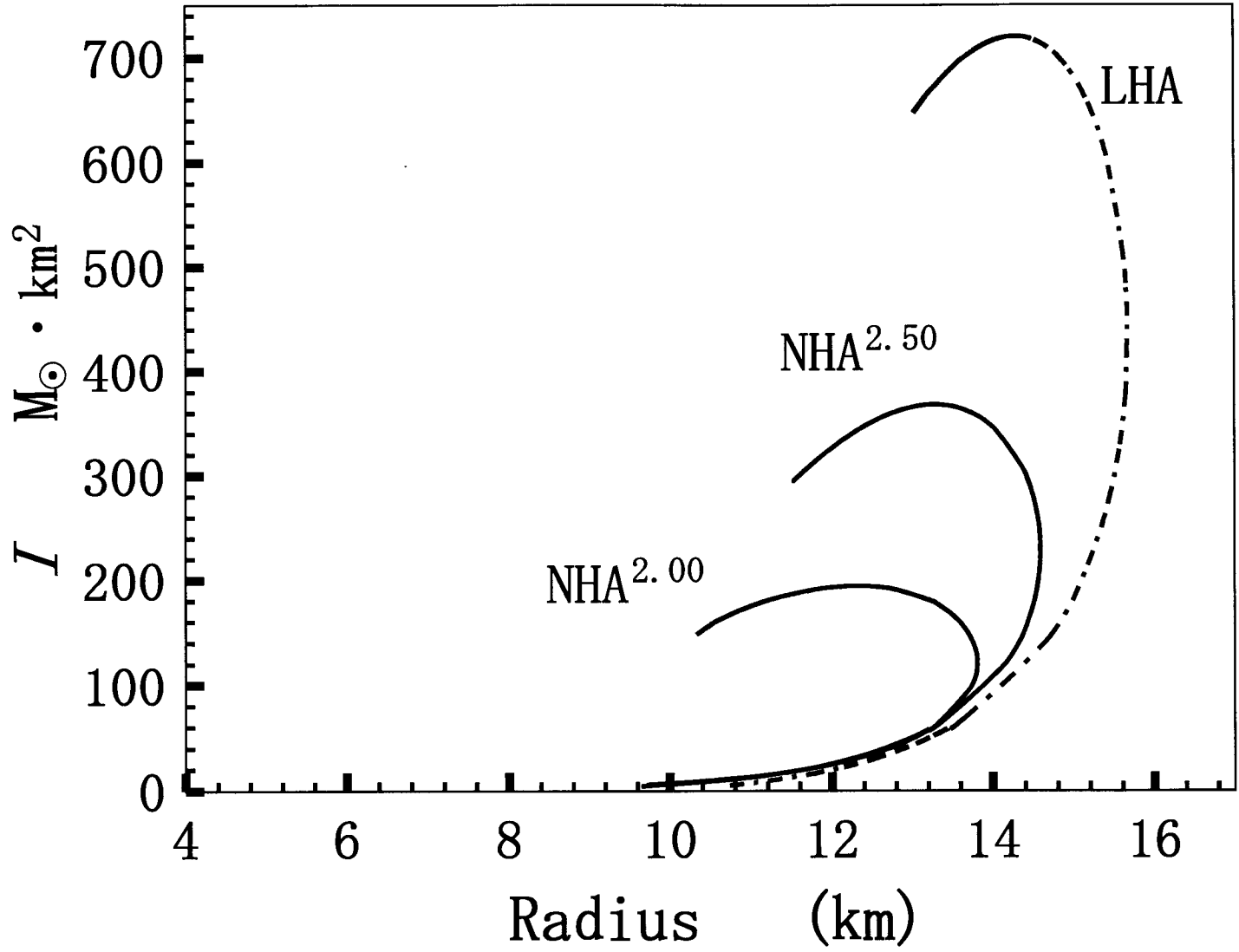
\item[Fig.~9.] The moments of inertia and radii of neutron stars.

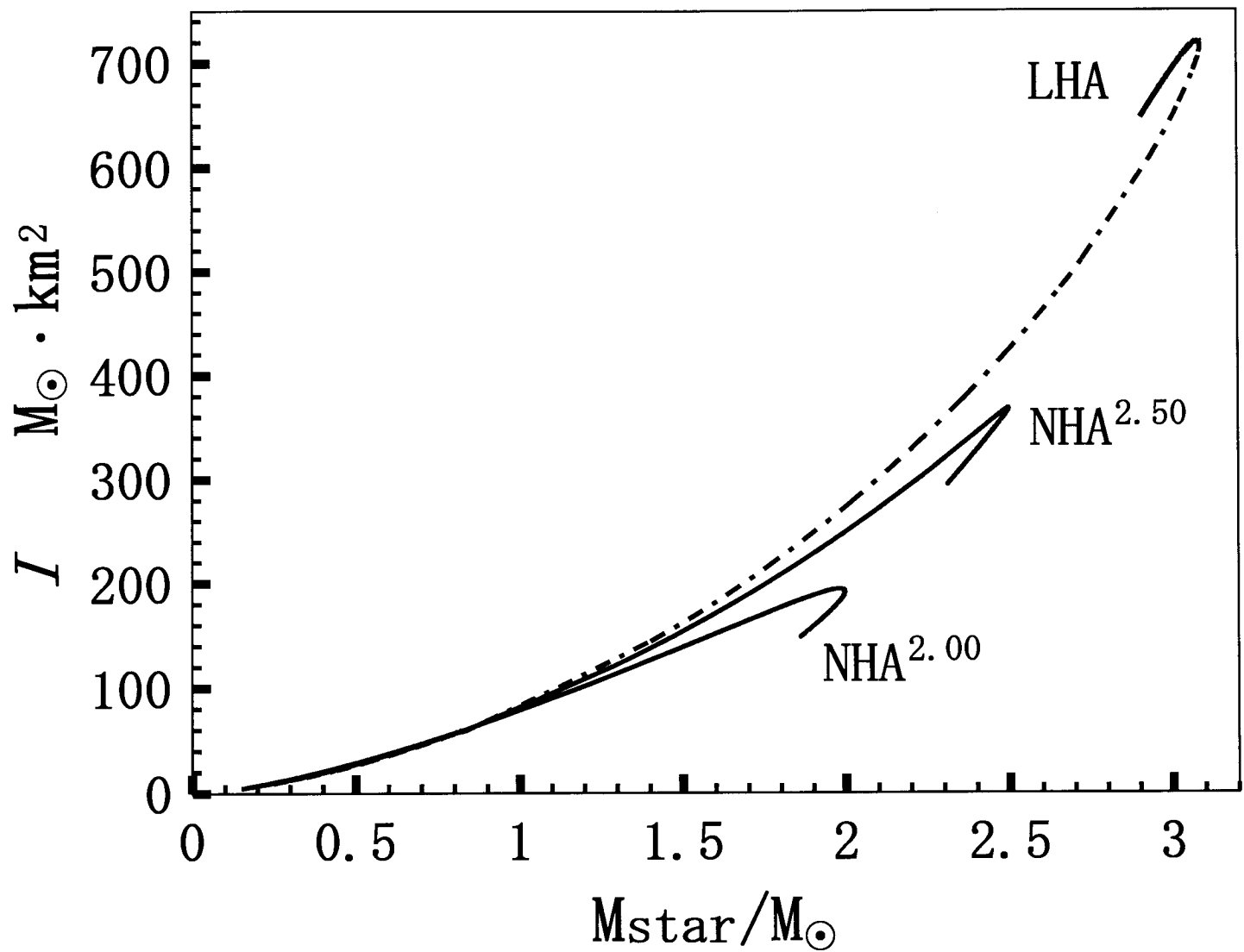
\item[Fig.~10.] The moments of inertia and masses of neutron stars.

\end{description}

\vspace{1.0cm}

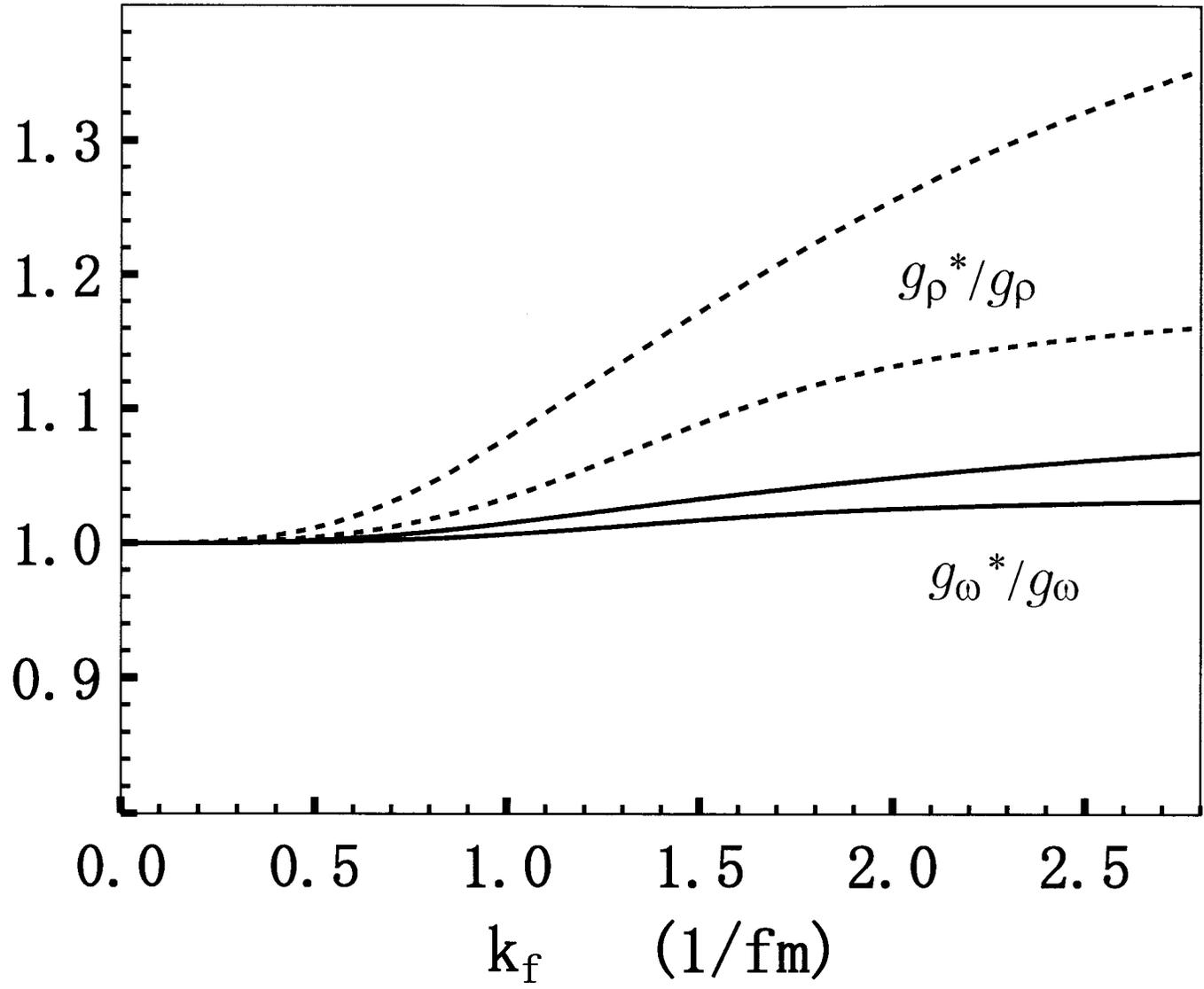
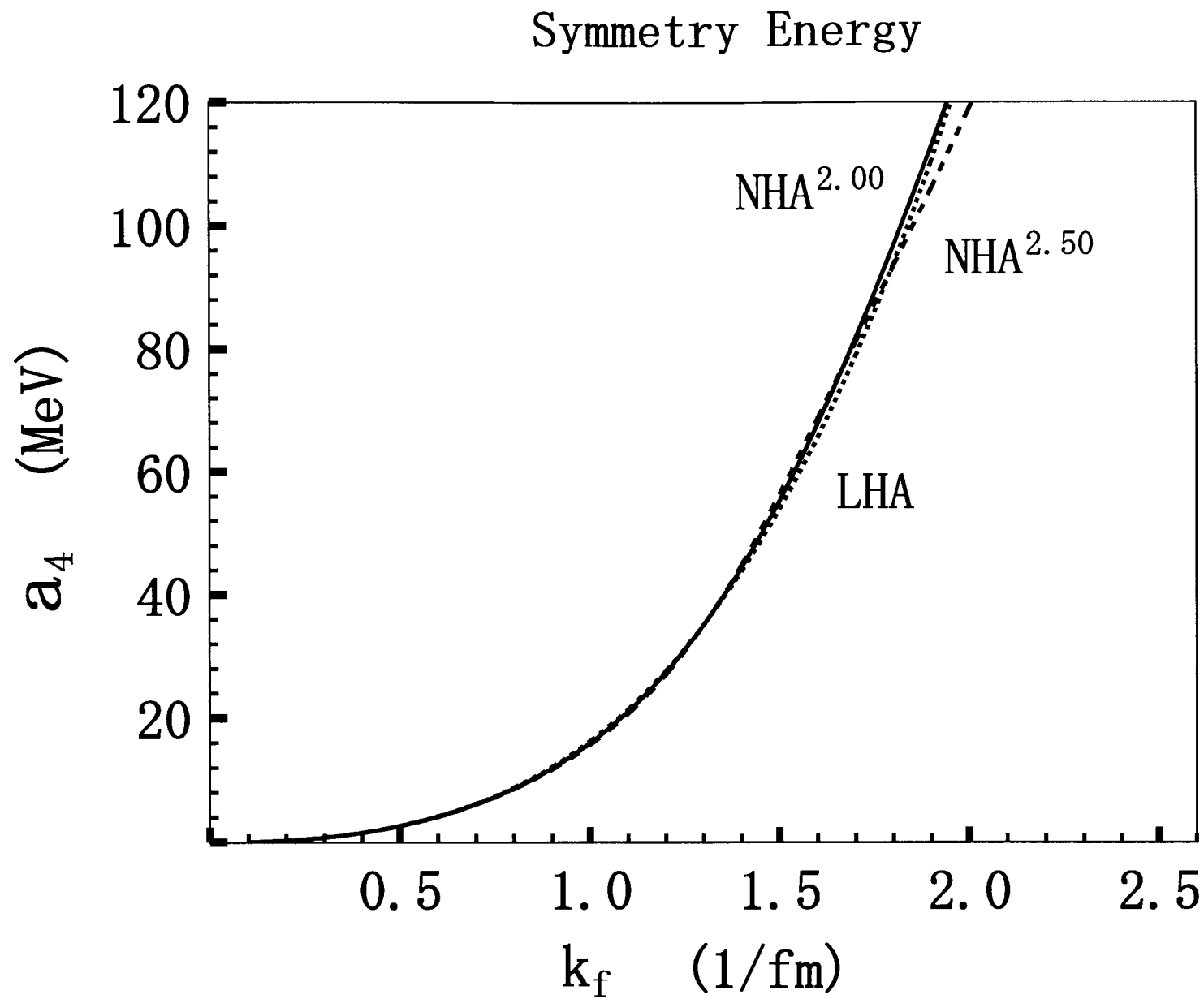
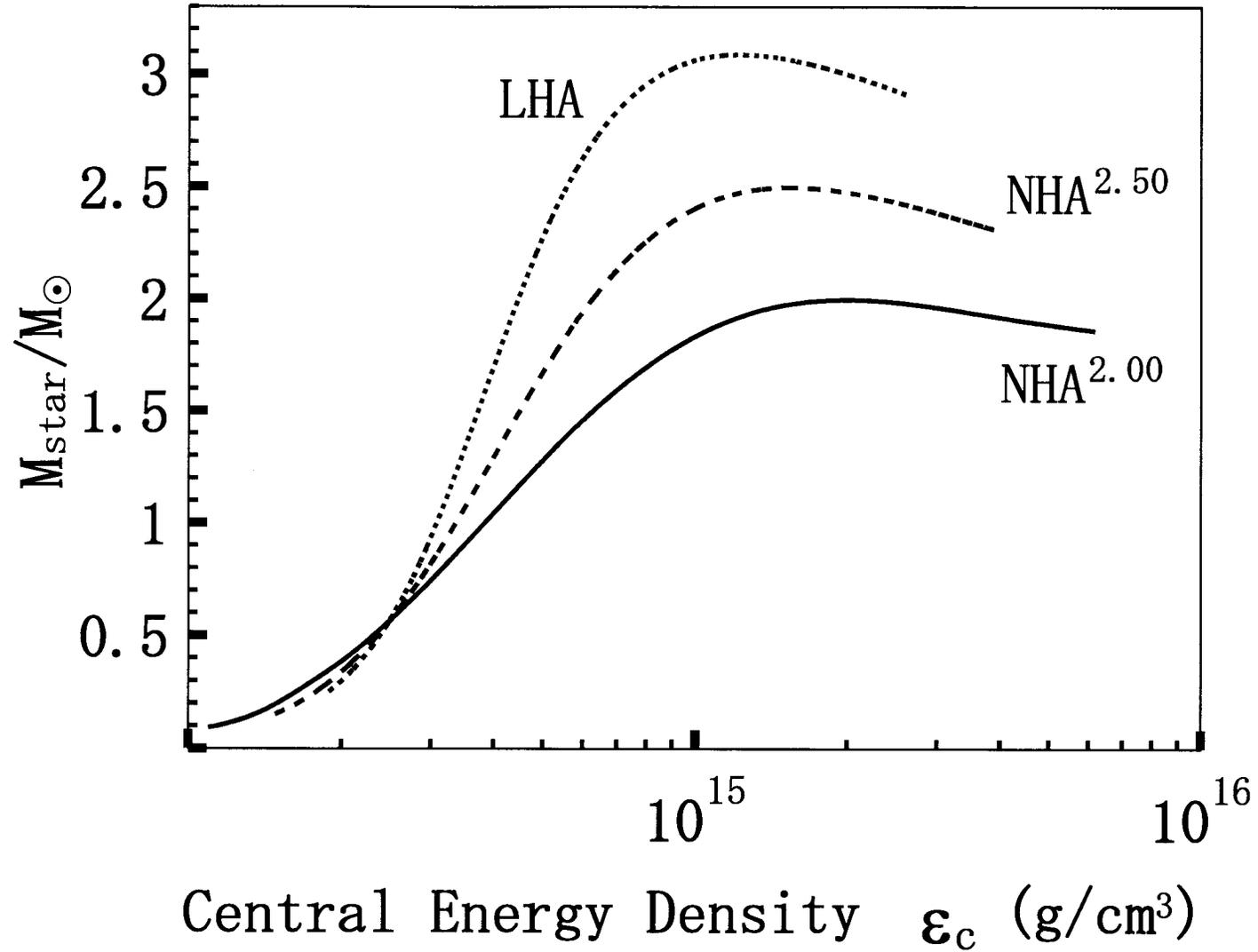
\end{document}